# Culsans: An Efficient Snoop-based Coherency Unit for the CVA6 Open Source RISC-V application processor

Riccardo Tedeschi
DEI
University of Bologna
Bologna, Italy
riccardo.tedeschi6@unibo.it

Luca Valente
Gianmarco Ottavi
DEI
University of Bologna
Bologna, Italy
{gianmarco.ottavi2,
luca.valente}@unibo.it

Enrico Zelioli
Nils Wistoff
IIS
ETH Zurich
Zurich, Switzerland
{ezelioli, nwistoff}@iis.ee.ethz.ch

Massimiliano Giacometti
Abdul Basit Sajjad
PlanV Tech
Munich, Germany
{massimiliano.giacometti,
abdul.basit}@planv.tech

Luca Benini
IIS, DEI
ETH Zurich, University of Bologna
Zurich, Switzerland
lbenini@iis.ee.ethz.ch

Davide Rossi
DEI
University of Bologna
Bologna, Italy
davide.rossi@unibo.it

*Abstract*— **Symmetric Multi-Processing (SMP) based on cache coherency is crucial for high-end embedded systems like automotive applications. RISC-V is gaining traction, and open-source hardware (OSH) platforms offer solutions to issues such as IP costs and vendor dependency. Existing multi-core cache-coherent RISC-V platforms are complex and not efficient for small embedded core clusters. We propose an open-source SystemVerilog implementation of a lightweight snoop-based cache-coherent cluster of Linux-capable CVA6 cores. Our design uses the MOESI protocol via the Arm's AMBA ACE protocol. Evaluated with Splash-3 benchmarks, our solution shows up to 32.87% faster performance in a dual-core setup and an average improvement of 15.8% over OpenPiton. Synthesized using GF 22nm FDSOI technology, the Cache Coherency Unit occupies only 1.6% of the system area.**

*Keywords- cache coherency; RISC-V; tightly coupled; CVA6; Culsans; ACE;*

## I. INTRODUCTION

Symmetric Multi-Processing based on Cache Coherency is critical for computing platforms in high-end embedded systems, such as those used in automotive applications. In this field, RISC-V is rapidly gaining acceptance, and Open-Source Hardware (OSH) platforms based on RISC-V have great potential for overcoming several issues, such as IP cost barriers, supply chain constraints, vendor captivity concerns, and non-recurring engineering (NRE) costs.

Current multi-core cache-coherent open-source RISC-V platforms use custom on-chip communication protocols and automated HDL generation, complicating the integration into third-party Systems on Chip (SoCs). Research platforms like OpenPiton [1] and ESP [2] use directory-based coherence to scale to many cores (> 4). However, the complexity of a distributed directory-based system is overkill for small embedded core clusters, leading to inefficiencies and area overheads. Rocket [3] offers a tightly coupled solution but relies on the Chisel hardware construction language to generate the HDL description, making it hard to develop a verification and integration strategy for SoCs where most of the IPs and system interconnect are based on industry-standard HDLs (e.g., SystemVerilog).

Thus, an open Cache Coherency Unit (CCU) designed for low overhead and high efficiency with a small core count (2-4), easy integration into custom SoCs, and full compatibility with commercial EDA flows has yet to be released. To close this gap, we propose Culsans, an open-source SystemVerilog implementation of a lightweight snoop-based tightly-coupled cache-coherent cluster of Linux-capable CVA6 cores [4], and we demonstrate its integration into Cheshire [5], an open RISC-V platform for domain-specific accelerators plug-in. Our solution implements the MOESI cache coherency protocol via the industry-standard Arm's AMBA ACE protocol, which extends the AXI protocol already supported in CVA6 with additional signals and channels aimed at memory coherency.

This work was supported by the Italian National Centre for HPC, Big Data and Quantum Computing – HPC (CN00000013) and the Technology Innovation Institute, Secure Systems Research Center, Abu Dhabi, UAE, PO Box: 9639.

Table 1. MOESI and ACE states mapping to status flags

| MOESI | ACE | Valid | Shared | Dirty |
|---|---|---|---|---|
| Modified | UniqueDirty | 1 | 0 | 1 |
| Owned | SharedDirty | 1 | 1 | 1 |
| Exclusive | UniqueClean | 1 | 0 | 0 |
| Shared | SharedClean | 1 | 1 | 0 |
| Invalid | Invalid | 0 | X | X |

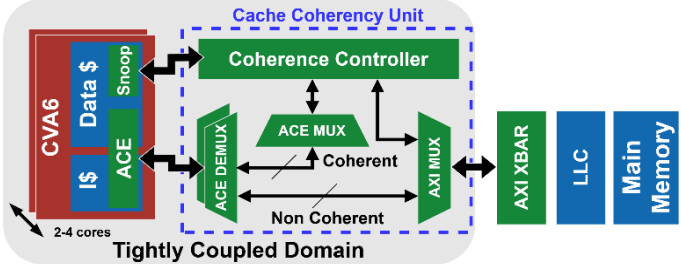

Figure 1. System Level view of the tightly coupled cluster of cores

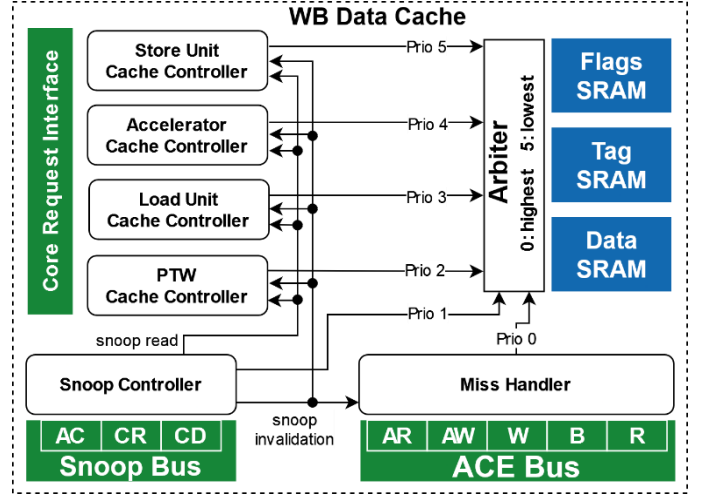

Figure 2. WB Data Cache Controllers with priority scheme and snoop control signals

A CCU was implemented to handle multiple outstanding memory requests in a pipelined fashion. The CVA6 Write-Back (WB) cache subsystem was updated to support AMBA ACE on top of AXI.

Our solution was evaluated using Splash-3 benchmarks [6] against OpenPiton. Under similar cache configurations, the tightly coupled cluster of cores proved to be up to 32.87% faster in a dual-core setup, with an average improvement of 15.8%. Moreover, the system was synthesized using GF 22nm FDSOI technology, and the area occupation of the Cache Coherent Unit amounted to only 1.6% of the overall system.

## II. ARCHITECTURE

### A. Memory Hierarchy

The proposed implementation of a snoop-based tightly coupled cache-coherent cluster of CVA6 cores is reported in Figure 1. The memory hierarchy features L1 Data and Instruction caches interacting with the CCU, while a Last Level Cache (LLC) shared among all the cores is inserted between the tightly coupled domain and the main memory.

### B. CVA6 Data Cache

On the core side, the pre-existing CVA6 WB data cache was extended to support ACE on top of AXI. The cache line status comprises three flags: valid, shared, and dirty. These additional flags are stored in the cache SRAM along with the cache line tag and data. The combination of the status three bits encodes the MOESI/ACE states, as shown in Table 1.

The WB cache comprises multiple controllers that handle the requests issued by the core, as shown in Figure 2. In particular, the Page Table Walker (PTW), the Load Unit, the Accelerator, and the Store Unit have a dedicated controller each. In addition, the Miss Handler is responsible for miss requests towards the next memory level, Atomic Memory Operations, cache flushes, and writeback operations since it acts as an initiator on the AXI interface of the core. The arbitration on the SRAM's single port read/write port is handled via a statically assigned priority. The Miss Handler has the highest precedence, followed by the PTW, the Load Unit, the Accelerator, and the Store Unit.

An additional Snoop Controller was added to the Cache Subsystem to handle transactions on the snoop bus, namely the snoop request channel AC, the snoop response channel CR, and the snoop data channel CD as defined in the ACE protocol. The other cores use this additional snoop interface to access and invalidate cache lines. The Snoop Controller was assigned a priority second only to the Miss Handler to ensure that cache line status updates needed to safeguard coherency are served before any request issued by the core. Additional snoop control signals are propagated to the Miss Handler and the other Cache Controllers to indicate an external invalidation or read request on a given cache line. A Cache Controller requesting unique access (i.e. shared flag equal to 0) to a cache line must ensure that no snoop read is performed concurrently, which indicates a transition to a shared condition of the cache line. Similarly, the Miss Handler must monitor the same event when fetching a cache line for unique access. In addition, a cache line might be invalidated by the Snoop Controller while a Cache Controller has ongoing operations on it, thus a snoop invalidation signal is used to propagate the information on the address being invalidated.

Lastly, the Miss Handler was updated to handle coherent and non-coherent requests via the fields added to the traditional AXI channels to encode the ACE defined transactions. This controller generates both data and invalidation requests towards the CCU.

### C. CVA6 Instruction Cache

The coherence of the Instruction Cache is required in specific applications where a core can generate instructions to be executed on a different core, such as in the Bao embedded hypervisor [7]. To accommodate this need, the Instruction Cache can be configured via an RTL-level parameter to generate coherent fetch requests and ensure coherency with the data caches of the clustered cores. Otherwise, this feature can be turned off to avoid additional snoop traffic.

*D. Cache Coherency Unit*

Figure 1 shows the interconnect architecture of the CCU, which was developed as a completely new IP. The memory operations issued by the cores are routed depending on whether they are non-coherent or coherent by the ACE DEMUX block. In the first situation, the request is directly forwarded to the memory interface of the CCU. In the second case, the coherent requests issued by different cores are ordered in the ACE MUX depending on the arrival time according to a round-robin policy and are processed serially by the Coherence Controller, which is also the initiator of the Snoop Bus. Both coherent and non-coherent requests are eventually serialized towards the system crossbar by the AXI MUX block.

The internal organization of the controller is depicted in Figure 3. Three main blocks are present: Decoder, Memory Unit, and Snoop Unit. The Decoder processes the initiator core's write (AW channel) and read (AR channel) requests. Starting from the request address and the ACE coherency transaction issued by the initiator core, the Decoder generates the appropriate snoop transaction towards the remaining cores through the snoop request (AC) and response (CR) channels. The generation of the AC request and the processing of the ensuing CR response are decoupled and mutually nonblocking; thus, a new AC request can be generated without waiting for the previous CR response. A FIFO keeps track of the response order since no transaction ID is associated with the Snoop channels.

Suppose a data snoop is triggered, and one or more cores can provide the needed cache line. In that case, the first responder passes the data to the CCU via the CD data channel, and the response is buffered in the Snoop Unit, which forwards it to the initiator core by generating a burst response on the read response R channel. Similarly, if a snooped core issues a write-back, the CD data is stored in a FIFO inside the memory unit for later processing. The memory unit handles the memory interface of the CCU, either by serving memory operations generated by the initiator core or by generating AXI transactions once a snooped core issues a write-back.

The requests are moved across the different blocks in a pipelined fashion, and control flow is ensured via asynchronous handshakes between the Snoop Unit, the Memory Unit, and the Decoder. Moreover, the inherent channel parallelism of the AXI is leveraged to decouple requests from responses. Serialization of requests is enforced only when multiple requests are targeted to the same cache line. An associative table is used as a Collision Checker to keep track of currently accessed cache lines, and stalling happens if a collision is detected upon a lookup by the Decoder.

### III. RESULTS

We evaluated our solution using Splash-3 benchmarks, comparing its performance to OpenPiton in a dual-core setup. Both are based on the CVA6 core, but in our implementation the CVA6 core employs a WB data cache, while in OpenPiton it uses a Write Through (WT) configuration. As shown in Figure 4, our solution is up to 32.87% faster in a dual-core setup, with an average improvement of 15.8%. These results are mainly due to our solution's tightly coupled design, which incurs less latency than directory-based platforms intended for many cores.

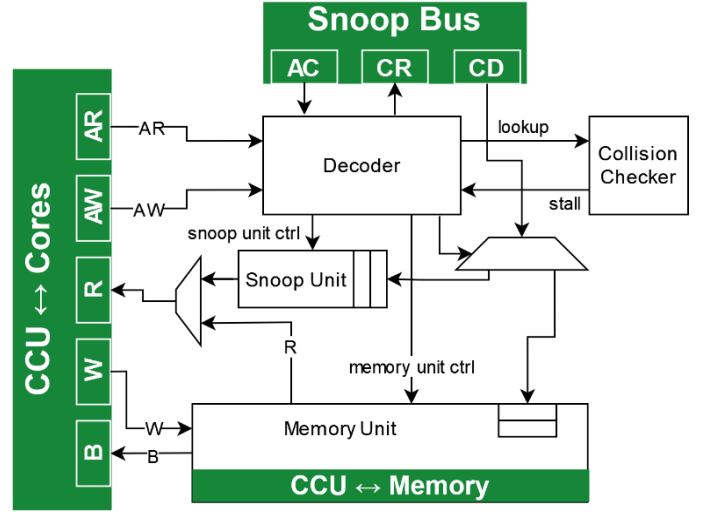

Figure 3. Internal block diagram of the Coherence Controller

The performance advantage of the CCU varies depending on the benchmark. Table 2 reports the profiling of pipeline stalls and memory operations normalized to the total number of instructions across the different benchmarks. Specific tests, namely FFT and RADIX, do not show improved performance with our snoop-based approach compared to the directory-based OpenPiton because of both their significant number of stalls and fewer memory operations than the other benchmarks. Thus, numerous operand-related pipeline stalls occur, and no advantage ensues from faster coherence transactions. On the contrary, benchmarks such as OCEAN or LU NC are characterized by a significant number of stalls along with a higher number of memory operations and benefit from the proposed changes.

In a first attempt to profile WB and WT caches implemented in CVA6 in a single core setup, we observed that the WB appears to be less performing than the WT, despite expecting from a theoretical point of view a performance advantage of the WB policy over the WT one. A possible explanation stems from the observation that several implementation bottlenecks are present in the available WB cache, leading to sub-optimal handling of transactions. Moreover, the CVA6 core has limited support for multiple outstanding transactions.

We synthesized the design in GF 22nm FDSOI technology using topographical synthesis to assess the area overhead of the additional coherence logic. The CCU area occupation is 1.6% of the total design area, and the coherence logic does not limit the multi-core cluster maximum frequency.

Table 2. Stalls and Memory Operations normalized to the total number of instructions on Splash-3 benchmarks

| Benchmark | Ocean | Barnes | Chol. | Rad. | FMM | Wat. nsqrd | LU NC | Wat. Spatial | LU Cont | FFT | Radix |
|---|---|---|---|---|---|---|---|---|---|---|---|
| **Stalls/Instr** | 3.05 | 0.67 | 1.07 | 0.48 | 0.60 | 0.90 | 3.66 | 0.69 | 0.97 | 1.50 | 2.35 |
| **Mem. Op/Instr** | 0.35 | 0.44 | 0.28 | 0.35 | 0.21 | 0.31 | 0.36 | 0.32 | 0.36 | 0.28 | 0.23 |

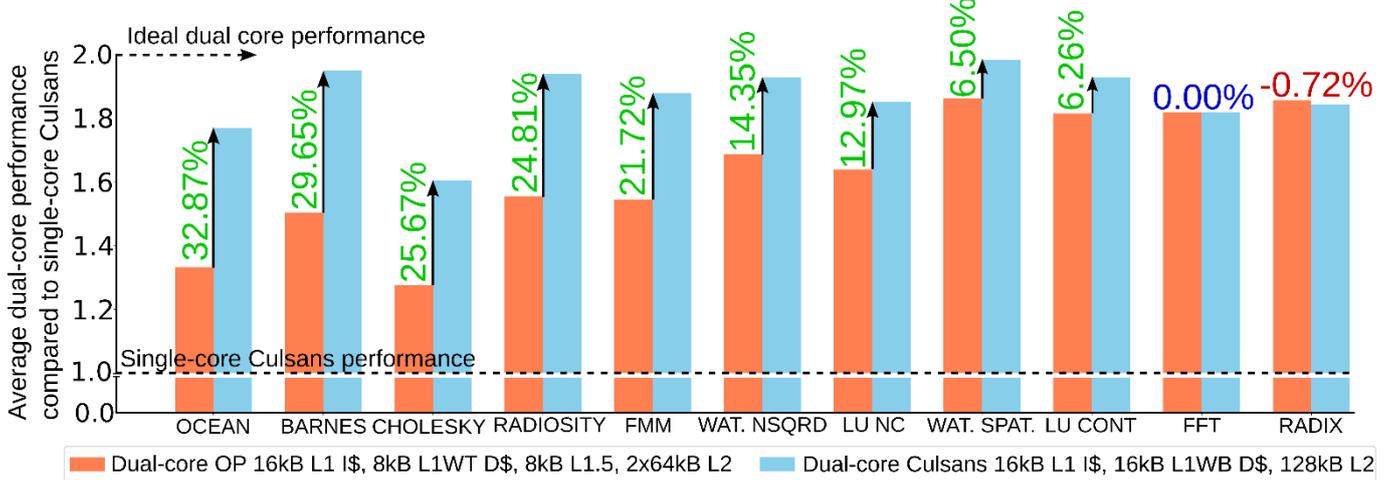

Figure 4. Performance comparison with respect to OpenPiton on the Splash-3 benchmarks

## IV. CONCLUSION

We presented Culsans[1], a snoop-based coherency unit for a tightly coupled cluster of CVA6 Open Source RISC-V application processors. The MOESI protocol is implemented via the industry-standard Arm's AMBA ACE protocol. The proposed architecture was fully integrated into the Cheshire platform and was evaluated against OpenPiton using the Splash-3 benchmarks. Our solution is up to 32.87% faster in a dual-core setup, with an average improvement of 15.8%. The area occupation of the new CCU is less than 2% of the entire dual-core system.

Future developments will focus on supporting advanced non-blocking and performance-oriented caches, such as the HPDCache [8], and more powerful cores, e.g. T-Head 910 [9], thanks to the use of the standardized AMBA ACE protocol. In addition, the Power, Performance, and Area analysis will be extended to larger clusters (4-8 cores). The hardware developed in this work is open-source to support an innovation ecosystem for high-performance, safety-critical embedded systems. Further work focusing on reliability and predictability in an embedded tightly coupled cluster of cores will be enabled by the availability of the proposed platform.

## REFERENCES

[1] J. Balkind, M. McKeown, Y. Fu, T. Nguyen, Y. Zhou, A. Lavrov, M. Shahrad, A. Fuchs, S. Payne, X. Liang and others, "OpenPiton: An open source manycore research framework," ACM SIGPLAN Notices, vol. 51, p. 217–232, 2016.

[2] P. Mantovani, D. Giri, G. Di Guglielmo, L. Piccolboni, J. Zuckerman, E. G. Cota, M. Petracca, C. Pilato and L. P. Carloni, "Agile SoC development with open ESP," in Proceedings of the 39th International Conference on Computer-Aided Design, 2020.

[3] K. Asanovic, R. Avizienis, J. Bachrach, S. Beamer, D. Biancolin, C. Celio, H. Cook, D. Dabbelt, J. Hauser, A. Izraelevitz and others, "The rocket chip generator," EECS Department, University of California, Berkeley, Tech. Rep. UCB/EECS-2016-17, vol. 4, p. 6–2, 2016.

[4] F. Zaruba and L. Benini, "The Cost of Application-Class Processing: Energy and Performance Analysis of a Linux-Ready 1.7-GHz 64-Bit RISC-V Core in 22-nm FDSOI Technology," IEEE Transactions on Very Large Scale Integration (VLSI) Systems, vol. 27, pp. 2629-2640, November 2019.

[5] A. Ottaviano, T. Benz, P. Scheffler and L. Benini, "Cheshire: A Lightweight, Linux-Capable RISC-V Host Platform for Domain-Specific Accelerator Plug-In," IEEE Transactions on Circuits and Systems II: Express Briefs, vol. 70, pp. 3777-3781, 2023.

[6] C. Sakalis, C. Leonardsson, S. Kaxiras and A. Ros, "Splash-3: A properly synchronized benchmark suite for contemporary research," in 2016 IEEE International Symposium on Performance Analysis of Systems and Software (ISPASS), 2016.

[7] J. Martins and S. Pinto, "Bao: A modern lightweight embedded hypervisor," in Proc. Embedded World Conf., 2020.

[8] C. Fuguet, "HPDcache: Open-source high-performance L1 data cache for RISC-V cores," in Proceedings of the 20th ACM International Conference on Computing Frontiers, 2023.

[9] C. Chen, X. Xiang, C. Liu, Y. Shang, R. Guo, D. Liu, Y. Lu, Z. Hao, J. Luo, Z. Chen, C. Li, Y. Pu, J. Meng, X. Yan, Y. Xie and X. Qi, "Xuantie-910: A Commercial Multi-Core 12-Stage Pipeline Out-of-Order 64-bit High Performance RISC-V Processor with Vector Extension : Industrial Product," in 2020 ACM/IEEE 47th Annual International Symposium on Computer Architecture (ISCA), 2020.

---

[1] Repository URL: https://github.com/pulp-platform/culsans